%% file: main.tex
\documentclass{INTERSPEECH2024}

\interspeechcameraready

\usepackage{multicol}
\usepackage{multirow}
\usepackage{color, colortbl}
\usepackage{xcolor}
\colorlet{eng}{blue!10}
\colorlet{cmn}{teal!10}
\colorlet{multi}{yellow!10}
\colorlet{euro}{orange!10}
\usepackage{pifont}
\usepackage{hyperref}
\usepackage{graphicx}
\usepackage{amsmath}
\usepackage{subcaption}
\newcommand{\cmark}{\ding{51}}%
\newcommand{\xmark}{\ding{55}}%

\title{Singing Voice Data Scaling-up: \\ An Introduction to ACE-Opencpop and ACE-KiSing}

\makeatletter
\def\thanks#1{\protected@xdef\@thanks{\@thanks
        \protect\footnotetext{#1}}}
\makeatother

\name[affiliation={1,*}]{Jiatong}{Shi}
\name[affiliation={2,*}]{Yueqian}{Lin}
\name[affiliation={3,4}]{Xinyi}{Bai}
\name[affiliation={}]{Keyi}{Zhang}
\name[affiliation={5}]{Yuning}{Wu}
\name[affiliation={5}]{Yuxun}{Tang}
\name[affiliation={6}]{Yifeng}{Yu} 
\name[affiliation={5,^\star}]{Qin}{Jin}
\name[affiliation={1}]{Shinji}{Watanabe}
\address{  \thanks{* Equal contribution.}  \thanks{ $^\star$ Corresponding Author.}
$^{1}$ Carnegie Mellon University,
        $^{2}$ Duke Kunshan University,
    $^{3}$ Cornell University,
	$^{4}$ Multimodal Art Projection Community, 
        $^{5}$ Renmin University of China,
        $^{6}$  Georgia Institute of Technology}
\email{jiatongs@cs.cmu.edu}

\usepackage[
backend=biber,
style=ieee,
citestyle=numeric-comp,
maxbibnames=10,
maxcitenames=20,
doi=false,isbn=false,url=false,eprint=false
]{biblatex}

\defbibheading{bibliography}[\refname]{}

\DeclareSourcemap{
	\maps[datatype=bibtex, overwrite=true]{
		\map{
			\step[fieldsource=booktitle,
			match=\regexp{.*Interspeech.*},
			replace={Proc. Interspeech}]
			\step[fieldsource=journal,
			match=\regexp{.*INTERSPEECH.*},
			replace={Proc. Interspeech}]
			\step[fieldsource=booktitle,
			match=\regexp{.*ICASSP.*},
			replace={Proc. ICASSP}]
			\step[fieldsource=booktitle,
			match=\regexp{.*icassp_inpress.*},
			replace={Proc. ICASSP (in press)}]
			\step[fieldsource=booktitle,
			match=\regexp{.*Acoustics,.*Speech.*and.*Signal.*Processing.*},
			replace={Proc. ICASSP}]
			\step[fieldsource=booktitle,
			match=\regexp{.*International.*Conference.*on.*Learning.*Representations.*},
			replace={Proc. ICLR}]
			\step[fieldsource=booktitle,
			match=\regexp{.*International.*Conference.*on.*Computational.*Linguistics.*},
			replace={Proc. COLING}]
			\step[fieldsource=booktitle,
			match=\regexp{.*SIGdial.*Meeting.*on.*Discourse.*and.*Dialogue.*},
			replace={Proc. SIGDIAL}]
			\step[fieldsource=booktitle,
			match=\regexp{.*International.*Conference.*on.*Machine.*Learning.*},
			replace={Proc. ICML}]
			\step[fieldsource=booktitle,
			match=\regexp{.*North.*American.*Chapter.*of.*the.*Association.*for.*Computational.*Linguistics:.*Human.*Language.*Technologies.*},
			replace={Proc. NAACL}]
			\step[fieldsource=booktitle,
			match=\regexp{.*Empirical.*Methods.*in.*Natural.*Language.*Processing.*},
			replace={Proc. EMNLP}]
			\step[fieldsource=booktitle,
			match=\regexp{.*Association.*for.*Computational.*Linguistics.*},
			replace={Proc. ACL}]
			\step[fieldsource=booktitle,
			match=\regexp{.*Automatic.*Speech.*Recognition.*and.*Understanding.*},
			replace={Proc. ASRU}]
			\step[fieldsource=booktitle,
			match=\regexp{.*Spoken.*Language.*Technology.*},
			replace={Proc. SLT}]
			\step[fieldsource=booktitle,
			match=\regexp{.*Speech.*Synthesis.*Workshop.*},
			replace={Proc. SSW}]
			\step[fieldsource=booktitle,
			match=\regexp{.*workshop.*on.*speech.*synthesis.*},
			replace={Proc. SSW}]
			\step[fieldsource=booktitle,
			match=\regexp{.*Advances.*in.*neural.*information.*processing.*},
			replace={Proc. NeurIPS}]
			\step[fieldsource=booktitle,
			match=\regexp{.*Advances.*in.*Neural.*Information.*Processing.*},
			replace={Proc. NeurIPS}]
			\step[fieldsource=booktitle,
			match=\regexp{.*Workshop.*on.* Applications.* of.* Signal.*Processing.*to.*Audio.*and.*Acoustics.*},
			replace={Proc. WASPAA}]
			\step[fieldsource=publisher,
			match=\regexp{.+},
			replace={{}}]
			\step[fieldsource=month,
			match=\regexp{.+},
			replace={{}}]
			\step[fieldsource=location,
			match=\regexp{.+},
			replace={{}}]
			\step[fieldsource=address,
			match=\regexp{.+},
			replace={{}}]
			\step[fieldsource=organization,
			match=\regexp{.+},
			replace={{}}]
		}
	}
}


\begin{document}

\maketitle
 
\begin{abstract}
In singing voice synthesis (SVS), generating singing voices from musical scores faces challenges due to limited data availability. This study proposes a unique strategy to address the data scarcity in SVS. We employ an existing singing voice synthesizer for data augmentation, complemented by detailed manual tuning, an approach not previously explored in data curation, to reduce instances of unnatural voice synthesis. This innovative method has led to the creation of two expansive singing voice datasets, ACE-Opencpop and ACE-KiSing, which are instrumental for large-scale, multi-singer voice synthesis. Through thorough experimentation, we establish that these datasets not only serve as new benchmarks for SVS but also enhance SVS performance on other singing voice datasets when used as supplementary resources. 
The corpora, pre-trained models, and their related training recipes are publicly available at ESPnet-Muskits (\url{https://github.com/espnet/espnet}).

\end{abstract}

\section{Introduction}
\label{sec: intro}

Singing voice synthesis (SVS) has emerged as an active research field within artificial intelligence-generated content~(AIGC). The task focuses on synthesizing singing voices from provided musical scores, which includes aligning lyrics, pitch, and note duration~\cite{cook1996singing, hono2021sinsy, kenmochi2007vocaloid}. In recent years, similar to text-to-speech (TTS), non-autoregressive singing voice synthesis based on sequence-to-sequence models has become the dominant approach~\cite{gu2020bytesing, lu2020xiaoicesing, seq2seqmodel, RNN, liu2022diffsinger, VISinger, visinger2}.

However, compared to TTS, SVS research often faces limited data availability due to strict copyright laws, the need for professional recording environments, and extensive annotation time. Various approaches to optimize training with limited datasets have been explored~\cite{RNN, guo2022singaug, ren2020deepsinger, choi2022melody, hong2023unisinger, xue2021learn2sing, shi2022muskits, zhou2023bisinger}. Some studies utilize additional data sources, such as crawled data or corpora combinations~\cite{ren2020deepsinger, xue2021learn2sing, choi2022melody, shi2022muskits, zhou2023bisinger}, others construct specialized loss functions for training stabilization~\cite{RNN}, and some apply data augmentation techniques for enhanced data diversity~\cite{guo2022singaug}.

In addition to algorithmic advancements, several benchmarks for SVS have been proposed to address data scarcity. Early works primarily focused on single-singer corpora~\cite{saino2006hmm, ogawa2021tohoku, wang22b_interspeech}.\footnote{Many databases are open-sourced without publications, such as \href{https://parapluie2c56m.wixsite.com/mysite}{Amaboshi Ciphyer Utagoe DB}, \href{https://ksdcm1ng.wixsite.com/njksofficial}{Yuuri Natsume Singing DB}, \href{https://sites.google.com/view/oftn-utagoedb/\%E3\%83\%9B\%E3\%83\%BC\%E3\%83\%A0}{Ofuton P Utagoe DB}, etc.} More recent efforts, such as M4Singer~\cite{zhang2022m4singer}, provide multi-singer support but with limited data per singer. In~\cite{shi2022muskits}, a multilingual-multi-singer corpus was compiled from six different sources, though this required additional fine-tuning on single-singer databases.

In practical applications, singing voices often undergo an editing process, including detailed manual tuning using digital signal processing techniques or parametric vocoders~\cite{langford2013digital, jackson2015digital}. As AI singing synthesizers have advanced, more amateur and professional musicians have started utilizing manual tuning to mitigate unnatural voices within musical signals. These tuning procedures offer valuable insights into the challenges of existing singing synthesis systems.



Building on the concept of manual voice editing, this paper introduces a novel data curation approach to create extensive singing voice corpora for SVS research. We employ manual tuning to mitigate unnatural voice outputs from the synthesizer, a method that has not been extensively investigated in previous studies. Utilizing \href{https://ace-studio.timedomain.cn/}{ACE Studio}, a widely recognized singing synthesizer, we execute this data curation process on source singing data from Opencpop~\cite{wang22b_interspeech} and KiSing~\cite{shi2022muskits}. The resultant corpora, ACE-Opencpop and ACE-KiSing, offer about 130 hours and 32.5 hours of singing from 30 and 34 vocalists, respectively. Through detailed experimentation with various models, we demonstrate that these corpora can significantly advance SVS research, acting as substantial large-scale SVS benchmarks. They also serve as valuable resources for augmenting other singing voice datasets, showing effectiveness in both in-domain and out-of-domain scenarios. Both ACE-Opencpop and ACE-KiSing are available under the CC-BY-NC-4.0 license. The key contributions of our study include:
\begin{itemize}
    \item Introducing a unique data curation method that incorporates a singing synthesizer and manual tuning to develop new singing voice corpora for SVS research.
    \item Releasing two large-scale multi-singer singing voice databases, ACE-Opencpop and ACE-KiSing, established through our specialized data curation approach.\footnote{\url{https://huggingface.co/datasets/espnet/{kising-v2-segments,ace-opencpop-segments}}}
    \item Thorough experimental validation shows that the proposed databases can serve as new SVS benchmarks or augment existing SVS datasets by transfer learning or joint-training.
\end{itemize}

\input{tables/corpora}

\section{The Corpora}

\subsection{Curation process}
\label{ssec: curation process}

 Both ACE-Opencpop and ACE-KiSing are synthesized using ACE Studio with additional manual tuning. The data curation process encompasses several major stages:

\textbf{Data Preparation:} Preparing the source data is essential for the data curation process, which involves preparing the singing voice of various artists and the corresponding musical scores. The musical score includes elements such as the sequence of musical notes, note durations, and syllable assignments for each note. Our approach also incorporates phone alignment of source singing voices within each syllable.


\textbf{Information Verification and Correction:} This stage involves enlisting individuals with musical expertise to meticulously refine the music score, typically in MIDI format, ensuring it accurately aligns with the singing voice. These experts rectify any discrepancies found in the initial phone alignment of the source singing voice as well.

\textbf{Tuning for Voice Match:} This phase involves an iterative process of editing the synthesized voice to achieve an optimal match with the source voice, which contains several steps: selecting a virtual singer whose voice closely resembles the source, modifying the fundamental frequency (F0) contour, incorporating additional breathing sounds, adjusting the vibrato, and fine-tuning the syllable duration. These modifications aim to ensure that the synthesized voice faithfully replicates the original song in the source database.

\textbf{Tuning for Singer Adaptation:} This process is repeated for each singer included in our study. As the singing range is different across singers, we filter out phrases that include unnatural segments due to the singing range mismatch. For ACE-Opencpop, we feature 30 singers, whereas ACE-KiSing includes 33. All voices are generated using ACE Studio version 1.7.2, and the data format is consistent across both corpora: 48kHz, single channel, 16-bit PCM.

\subsection{ACE-Opencpop}


The ACE-Opencpop corpus is derived from the 5.2-hour Opencpop benchmark~\cite{wang22b_interspeech} and inherits its song list. We employ the MIDI music scores from the benchmark, adhering to the data curation stages described in Section~\ref{ssec: curation process}. The voices in ACE-Opencpop are presented at song level (each song has a corresponding waveform), with the capability for segmentation aligning with the original Opencpop dataset's configuration. Owing to the utilization of identical music scores, ACE-Opencpop exhibits the same statistical characteristics as the original Opencpop, including song duration, pitch distribution, and phonetic (phone) distribution.

\subsection{ACE-KiSing}
\label{ssec: kising}

ACE-KiSing, an extension of the 2021 KiSing database \cite{shi2022muskits}, incorporates eight additional songs by singer Kiki Zhang, with corresponding music scores. Notably, this expansion introduces four English songs, rendering ACE-KiSing a bilingual singing voice corpus. Although bilingual singing synthesis is not the primary focus of this study, these data are included in our public release. A distinctive feature of ACE-KiSing, in comparison to previous works, is its emphasis on frequent melisma\footnote{Melisma refers to the technique in singing where a single syllable is extended over a sequence of notes.} techniques. Approximately 27.0\% of ACE-KiSing's singing phrases exhibit melisma, a significant increase from that 1.5\% observed in Opencpop.

\subsection{Comparison to other corpora}

As highlighted in Section~\ref{sec: intro}, the field of singing voice processing has witnessed several significant contributions in open-source benchmarks. In Table~\ref{tab:dataset}, we summarize some representative corpora that incorporate singing voices in their collections. 

A common limitation is that many databases featuring singing voices either lack explicit music score information or offer insufficient data per singer for robust SVS. These datasets are predominantly tailored for tasks like music information retrieval, singing voice conversion, or singing-focused vocoding~\cite{hsu2010mir, duan2013nus, SHARMA20219, tamaru2020jvs, huang21multisinger}. In SVS contexts, previous studies have mitigated these limitations by extracting pitch directly from the singing voice and using these pitch sequences as surrogate music score information \cite{liu2022diffsinger}. However, this method inherently requires a reference singing voice for the synthesis process, aligning it more closely with singing voice conversion applications.

Databases that do provide comprehensive music score information are typically limited to single-singer recordings \cite{liu2022diffsinger, wang22b_interspeech, shi2022muskits, jsutsong, ogawa2021tohoku, saino2006hmm}. In contrast, M4Singer \cite{zhang2022m4singer} and SingStyle111 \cite{dai2023singstyle111} feature 20 and 8 singers, respectively, but offer less than 2 hours of singing voice per artist.

In comparison, ACE-Opencpop and ACE-KiSing stand out with their coverage of approximately 30 singers each, representing the 3rd and 4th largest diversity of singers in SVS corpora, respectively. Moreover, ACE-Opencpop offers more than 100 hours of data, which is conducive to large-scale singing synthesis research. ACE-KiSing is particularly noteworthy for its focus on the melisma technique, offering a unique dimension to singing voice studies.

\section{Experiments}

\subsection{Experimental tasks}
\label{ssec: exp task}

To fully investigate the proposed corpora, we conduct three tasks in this work, including:

\begin{itemize}
    \item \textbf{Direct SVS}:  This task involves direct SVS using the proposed corpora, where we evaluate the performance of different SVS models. We opt for two emblematic models from ESPnet-Muskits~\cite{shi2022muskits}, namely Xiaoice~\cite{lu2020xiaoicesing} and VISinger2~\cite{visinger2}. These models represent the two-stage SVS model (Xiaoice, comprising an acoustic model and a vocoder) and the end-to-end SVS model (VISinger2), respectively. Given VISinger2's superior performance comapred to Xiaoice, it is selected for subsequent tasks.
    \item \textbf{Transfer learning}: This task leverages the pre-trained model from ACE-Opencpop, utilizing its substantial data volume (around 130 hours), to implement transfer learning on other smaller-scale singing synthesis benchmarks. We evaluate this in two contexts: an in-domain transfer (ACE-Opencpop to Opencpop) and an out-of-domain transfer (ACE-Opencpop to Kiritan). The latter presents a significant domain shift, with ACE-Opencpop centered on Mandarin Pop songs and Kiritan on Japanese Animation songs. This task also includes a comparative analysis with SVS systems trained exclusively on the original datasets.
    \item \textbf{Joint training}: In the joint training task, we merge the proposed corpus with other established SVS benchmarks, using it as an augmentation resource to enrich the training data. The original KiSing study~\cite{shi2022muskits} highlighted difficulties in synthesizing high-quality singing voices with the database. This is due in particular to the prevalent use of the melisma technique, which is also examined in Section~\ref{ssec: kising}. In this research, we aim to assess how our data curation approach for KiSing (i.e., the generated corpus, ACE-KiSing) can address these challenges. By incorporating ACE-KiSing for supplementary training alongside the original KiSing corpus, we intend to enhance the model's learning experience and improve its performance. The effectiveness of this task is evaluated using the original KiSing test set, providing a measure of how our proposed corpus influences the quality of SVS (especially in contexts where complex vocal techniques like melisma are involved).
    
\end{itemize}

\subsection{Experimental model details}
\label{ssec: experiment model}

As mentioned in Section~\ref{ssec: exp task}, we use Xiaoice~\cite{lu2020xiaoicesing} and VISinger2~\cite{visinger2} for the experiments:

\begin{itemize}
    \item \textbf{Xiaoice}: The Xiaoice model follows the XiaoiceSing2 \cite{chunhui23_interspeech} by using conformer blocks \cite{gulati20_interspeech, guo2021recent} in both the encoder and decoder. Additionally, voice/unvoiced and log F0 prediction losses are added along with the Mel spectrogram prediction. For multi-singer SVS, the singer ID embedding is repeated along the time axis and concatenated with the hidden states before feeding to the decoder. The training and network hyperparameters align with the Xiaoice configuration detailed in the Opencpop recipe.\footnote{\scriptsize{\url{https://github.com/espnet/espnet/blob/master/egs2/opencpop/svs1}}} A HiFi-GAN vocoder, pre-trained for 250k steps on the relevant corpora, is used for voice generation, with parameters sourced from the ParallelWaveGAN repository.\footnote{\scriptsize{\url{https://github.com/kan-bayashi/ParallelWaveGAN/blob/master/egs/opencpop/voc1/conf/hifigan.v1.yaml}}}
    
    \item \textbf{VISinger2}:  This model \cite{visinger2} advances the VISinger \cite{VISinger} by incorporating a differentiable digital signal processing (DDSP) vocoder. Mirroring the VITS model \cite{kim2021conditional} in TTS, VISinger2 uses a conditional variational autoencoder, which is trained alongside adversarial objectives. It processes musical score input and employs a frame prior network with pitch prediction to enhance the flow learning process. The configuration for VISinger2 adheres to the specifications in ESPnet-Muskits. To facilitate multi-singer SVS, singer ID embeddings are integrated as global information within the VITS framework, affecting both the generator and discriminators.
\end{itemize}

To evaluate the models in different sampling rates, we conduct cascaded acoustic models with the vocoder (i.e., Xiaoice) in 24kHz and 44.1kHz for end-to-end SVS systems (i.e., VISinger2).

\begin{table}[t!]
  \caption{Direct SVS of ACE-Opencpop. 95\% confidence intervals of MOS are presented in parentheses. G.T.\textsuperscript{\#} are test samples in ACE-Opencpop test set. Source G.T. stands for the same singing phrases from the original Opencpop test set. }
  \vspace{-5pt}
  \label{tab:direct-opencpop}
  \centering
  \resizebox{\linewidth}{!}{%
  \begin{tabular}{l|ccccc}
    \toprule
    \textbf{Model}  & \textbf{MCD $\downarrow$} & \textbf{S. Acc. $\uparrow$} & \textbf{F0 RMSE $\downarrow$} & \textbf{SECS $\uparrow$} & \textbf{MOS $\uparrow$} \\
    \midrule
    Xiaoice  & 5.81 & 64.33 & 0.162 & 0.75 & 3.53 ($\pm$ 0.05) \\
    VISinger2 & \textbf{5.08} & \textbf{65.19} & \textbf{0.140} & \textbf{0.78} & \textbf{3.81} ($\pm$ 0.05) \\
    \midrule
    G.T.\textsuperscript{\#} & - & - & - & - & 4.35 ($\pm$ 0.06) \\
    Source G.T. & - & - & - & - & 4.69 ($\pm$ 0.06) \\
    \bottomrule
  \end{tabular}%
  \vspace{-15pt}
  }
\end{table}

\begin{table}[t!]
  \caption{Direct SVS of ACE-KiSing. 95\% confidence intervals of MOS are presented in parentheses. G.T.\textsuperscript{\#} are test samples in ACE-KiSing test set. Source G.T. are the same singing phrases from original KiSing test set.}
  \vspace{-5pt}
  \label{tab:direct-kising}
  \centering
  \resizebox{\linewidth}{!}{%
  \begin{tabular}{l|ccccc}
    \toprule
    \textbf{Model}  & \textbf{MCD $\downarrow$} & \textbf{S. Acc. $\uparrow$} & \textbf{F0 RMSE $\downarrow$} & \textbf{SECS $\uparrow$} & \textbf{MOS $\uparrow$} \\
    \midrule
    Xiaoice  & 6.10 & 62.93 & 0.199 & 0.77 & 3.29 ($\pm$ 0.06) \\
    VISinger2 & \textbf{5.24} & \textbf{64.50} & \textbf{0.185} & \textbf{0.80} & \textbf{3.64} ($\pm$ 0.06) \\
    \midrule
    G.T.\textsuperscript{\#} & - & - & - & - & 4.49 ($\pm$ 0.05) \\
    Source G.T.  & - & - & - & - &  4.51 ($\pm$ 0.07) \\
    \bottomrule
  \end{tabular}%
  }
  \vspace{-15pt}
\end{table}

\subsection{Evaluation metrics}

In our study, we conduct both objective and subjective evaluations on the synthesized singing voice samples. For the objective assessment, we utilize Mel cepstral distortion (MCD), semitone accuracy (S. Acc.), and logarithmic F0 root mean square error (F0 RMSE) as our metrics, consistent with the standards in previous research \cite{shi2022muskits, guo2022singaug, wu2023phoneix, wu2023systematic}. Given that the two SVS models employed have different output sampling rates, we resample all generated singing voices to 24kHz for the above metrics calculation to ensure a fair comparison. For direct SVS experiments, which support multi-singer synthesis, we introduce an additional objective metric, speaker embedding cosine similarity (SECS), using a pre-trained Rawnet3-based speaker embedding extractor \cite{jung20c_interspeech} provided by ESPnet-SPK \cite{jung2024espnet}.

In the subjective evaluation, we conduct a Mean Opinion Score (MOS) test to evaluate the perceptual quality of the synthesized voices. We randomly select 30 synthesized samples from each system for this test. These samples are then evaluated by 30 listeners using a 5-point scale, where 1 indicates ``unreasonable singing" and 5 signifies ``natural singing comparable to human performance". For all the experiments, we include ground truth (G.T.) MOS for comparison.  In Direct SVS, the G.T. from our corpora (ACE-Opencpop and ACE-KiSing) is labeled as G.T.\textsuperscript{\#}. Additionally, for reference, we provide MOS results for the source audios from the original Opencpop and KiSing databases.

\begin{table}[t!]
  \caption{In-domain transfer learning experiments with Opencpop. * means using  pre-trained ACE-Opencpop models. 95\% confidence intervals of MOS are presented in parentheses.}
  \vspace{-5pt}
  \label{tab:transfer-opencpop}
  \centering
  \resizebox{\linewidth}{!}{%
  \begin{tabular}{l|cccc}
    \toprule
    \textbf{Model}  & \textbf{MCD $\downarrow$} & \textbf{S. Acc. $\uparrow$} & \textbf{F0 RMSE $\downarrow$} & \textbf{MOS $\uparrow$} \\
    \midrule
    Xiaoice  & 9.26 & 59.22 & 0.185 & 2.78 ($\pm$ 0.03) \\
    Xiaoice* & 8.78 & 60.72 & 0.182 & 3.08 ($\pm$ 0.05) \\
    \midrule
    VISinger2 & 7.54 & 64.50 & 0.172 & 3.63 ($\pm$ 0.07) \\
    VISinger2* & \textbf{7.26} & \textbf{65.18} & \textbf{0.162} & \textbf{3.66} ($\pm$ 0.06) \\
    \midrule
    G.T. & - & - & - & 4.69 ($\pm$ 0.06) \\
    \bottomrule
  \end{tabular}%
  }
  \vspace{-5pt}
\end{table}

\begin{table}[t!]
  \caption{Out-domain transfer learning experiments with Kiritan-singing. * indicates the use of pre-trained ACE-Opencpop models. 95\% confidence intervals of MOS are presented in parentheses.}
  \vspace{-5pt}
  \label{tab:transfer-kiritan}
  \centering
  \resizebox{\linewidth}{!}{%
  \begin{tabular}{l|cccc}
    \toprule
    \textbf{Model}  & \textbf{MCD $\downarrow$} & \textbf{S. Acc. $\uparrow$} & \textbf{F0 RMSE $\downarrow$} & \textbf{MOS $\uparrow$} \\
    \midrule
    VISinger2 & \textbf{7.47} & 49.33 & 0.123 & 3.58 ($\pm$ 0.07) \\
    VISinger2* & 7.54 & \textbf{50.88} & \textbf{0.122} & \textbf{3.68 ($\pm$ 0.07)}  \\
    \midrule
    G.T. & - & - & - & 4.57 ($\pm$ 0.07) \\
    \bottomrule
  \end{tabular}%
  }
    \vspace{-15pt}
\end{table}

\subsection{Results and discussion}

\noindent \textbf{Direct SVS}: The results displayed in Tables~\ref{tab:direct-opencpop} and~\ref{tab:direct-kising} highlight the performance of direct SVS using the two corpora. We observe noticeable differences in MOS between the proposed ACE-Opencpop and original Opencpop, while the prior still have a relatively-high MOS of 4.35. Meanwhile, we do not find a significant gap in MOS between ACE-KiSing and KiSing. As we compare the proposed copora with Xiaoice and VISinger2 (see Tables~\ref{tab:direct-opencpop} and~\ref{tab:direct-kising}) there is a significant disparity in the subjective score, suggesting the high quality of the two corpora. This finding indicates that ACE-Opencpop and ACE-KiSing can serve as valuable benchmarks for evaluating existing and future large-scale multi-singer SVS systems.

Moreover, VISinger2 consistently surpasses Xiaoice across both objective and subjective evaluations. This superior performance of VISinger2 demonstrates its effective generalization capabilities in the context of large-scale multi-singer SVS.

\noindent \textbf{Transfer learning}: Tables~\ref{tab:transfer-opencpop} and~\ref{tab:transfer-kiritan} display the results of the in-domain and out-domain transfer experiments with ACE-Opencpop, respectively. The data in Table~\ref{tab:transfer-opencpop} reveal that employing the pre-trained model from ACE-Opencpop enhances both objective and subjective performance metrics in SVS. 
Compared to two-stage approaches (i.e., Xiaoice), end-to-end methods like VISinger2 consistently yield superior results, confirming findings from previous studies \cite{visinger2}. Nevertheless, there still is a huge difference between the highest-quality synthesized voices and actual singing, suggesting that there is plenty of room for further advancements in SVS systems.

In the out-domain transfer learning experiments, as shown in Table~\ref{tab:transfer-kiritan}, the application of the pre-trained model does not result in improvements in MCD metrics. However, we note reasonable enhancements in S. Acc, F0 RMSE, and subjective MOS. Given the substantial domain disparity between the two corpora, the pre-trained model evidently aids in generating more natural singing voices, as evidenced by significant improvements in metrics other than MCD.

\noindent \textbf{Joint training}: Table~\ref{tab:transfer-kising} presents the results from the ACE-KiSing joint-training scenario with original KiSing. Echoing the ACE-Opencpop experiments, models joint-trained with augmented voices consistently perform better or comparably in both objective and subjective evaluations. While the objective metrics show less pronounced improvements in aspects like  F0 RMSE, the S. Acc. and MOS scores exhibit notable differences. This discrepancy may stem from the melisma technique prevalent in the ACE-KiSing dataset, which tends to impact subjective listening quality more significantly than objective scores in F0 RMSE.

\subsection{Further discussion}

Table~\ref{tab:transfer-m4singer} provides an objective evaluation of various pre-trained models. Notably, M4Singer, as cataloged in Table~\ref{tab:dataset}, was the most extensive SVS corpus available prior to the release of ACE-Opencpop. The results interestingly reveal that simply fine-tuning the model pre-trained on M4Singer does not uniformly enhance the objective metrics. In contrast, the model pre-trained on ACE-Opencpop consistently shows substantial improvements. This distinction underscores the significance of ACE-Opencpop, not just as a standalone corpus but as a valuable augmentation resource for SVS tasks.

\begin{table}[t!]
  \caption{Joint-training experiments of ACE-KiSing and KiSing. + means the model trained with both ACE-KiSing and KiSing. 95\% confidence intervals of MOS are presented in parentheses.}
  \label{tab:transfer-kising}
  \vspace{-5pt}
  \centering
  \resizebox{\linewidth}{!}{%
  \begin{tabular}{l|cccc}
    \toprule
    \textbf{Model}  & \textbf{MCD $\downarrow$} & \textbf{S. Acc. $\uparrow$} & \textbf{F0 RMSE $\downarrow$} & \textbf{MOS $\uparrow$} \\
    \midrule
    VISinger2 & 5.55 & 68.12 & \textbf{0.168} & 3.58 ($\pm$ 0.07)\\
    VISinger2+ & \textbf{4.99} & \textbf{72.51} & 0.170 & \textbf{3.83 ($\pm$ 0.10)} \\
    \midrule
    G.T. & - & - & - & 4.51 ($\pm$ 0.07) \\
    \bottomrule
  \end{tabular}%
  }
    \vspace{-5pt}
\end{table}

\begin{table}[t!]
  \caption{Comparison with models from M4Singer. * indicates the use of pre-trained ACE-Opencpop models and $^\triangle$ indicates the use of pre-trained M4Singer models.}
  \vspace{-5pt}
  \label{tab:transfer-m4singer}
  \centering
  \resizebox{0.8\linewidth}{!}{%
  \begin{tabular}{l|ccc}
    \toprule
    \textbf{Model}  & \textbf{MCD $\downarrow$} & \textbf{S. Acc. $\uparrow$} & \textbf{F0 RMSE $\downarrow$} \\
    \midrule
    VISinger2 & 7.54 & 64.50 & 0.172 \\
    VISinger2$^\triangle$ & 8.51 & 60.54 & 0.195 \\
    VISinger2* & \textbf{7.26} & \textbf{65.18} & \textbf{0.162} \\
    \bottomrule
  \end{tabular}%
  }
    \vspace{-15pt}
\end{table}

\section{Conclusion}
In this study, we introduce a novel data curation approach that merges existing singing synthesizers with manual tuning. Utilizing this method, we develop two new corpora, ACE-Opencpop and ACE-KiSing, aimed at advancing research in SVS. These corpora are enhanced using ACE-Studio combined with manual tuning, which narrows the subjective quality gap with actual singing voices. Through comprehensive experiments, we demonstrate the effectiveness of ACE-Opencpop and ACE-KiSing in various SVS contexts, including direct singing synthesis, transfer learning, and joint training.

While our current work only focuses on these two corpora, our goal is to apply this curation strategy more broadly, creating additional datasets. This expansion could provide a more diverse and comprehensive set of resources, fostering innovation and development in the domain of singing synthesis.

\section{Acknowledgements}

Experiments of this work used the Bridges2 system at PSC and Delta system at NCSA through allocations CIS210014 and IRI120008P from the ACCESS program, supported by NSF grants \#2138259, \#2138286, \#2138307, \#2137603, and \#2138296. This work was partially supported by the  National Natural Science Foundation of China (No. 62072462). We would like to thank Shengyuan Xu and Pengcheng Zhu for their support in the data license. When utilizing singing voices from commercial software (e.g., ACE Studio) as training data for SVS, it is crucial to be aware of potential legal and ethical issues.




\section{References}
{
\printbibliography
}

\end{document}

%% file: tables/corpora.tex
\begin{table*}[h]
    \centering
    \caption{Comparison between the proposed corpora and existing singing voice benchmarks. SVS and SVC denote whether the corresponding corpus can be used for synthesis or conversion purpose. NC denotes non-commercial, while ND denotes non-derivatives.}
    \vspace{-5pt}
    \resizebox{\textwidth}{!}{
    \begin{tabular}{l|cccccccccc} 
        \toprule
        Dataset & Year & Language & \# of Songs & Dur. (hour) & \# of Singers & Dur./Singer (hour)  & Score & SVS & SVC & License  \\
        \midrule
        NIT SONG070 F001 \cite{saino2006hmm} & 2015 & JP & 31 & 1.2 & 1 & 1.2  & \cmark & \cmark & \xmark & CC \\
        JSUT Song \cite{jsutsong} & 2018 & JP & 27 & 0.4 & 1 & 0.4 &  \cmark & \cmark & \xmark & CC-NC-ND \\
        Kiritan \cite{ogawa2021tohoku} & 2021 & JP & 50 & 1.0 & 1 & 1.0 &  \cmark & \cmark & \xmark & Research-only \\
        KiSing-v1 \cite{shi2022muskits} & 2021 &  CN & 14 & 0.7 & 1 & 0.7 &  \cmark & \cmark & \xmark & CC-NC \\
        Opencpop \cite{wang22b_interspeech} & 2022 & CN & 100 & 5.2 & 1 & 5.2 &  \cmark & \cmark & \xmark & CC-NC-ND \\
        PopCS \cite{liu2022diffsinger}& 2022 &  CN & 117 & 5.9 & 1 & 5.9 &  \xmark & \xmark & \xmark & CC-NC \\
        \midrule
        NUS-48E \cite{duan2013nus} & 2013 & EN & 48  & 1.9 & 12 & 0.2 &  \xmark & \xmark & \cmark &  Research-only \\
        NHSS \cite{SHARMA20219} & 2019 & EN & 100  & 4.8 & 10 & 0.5 &  \xmark & \xmark & \cmark &  Research-only \\
        JVS-MuSiC \cite{tamaru2020jvs} & 2020 & JP & 2 & 2.3 & 100 & 0.2 &  \xmark &  \xmark & \cmark & CC \\
        OpenSinger \cite{huang21multisinger} & 2021 & CN & 1146 & 50.0 & 66 & 0.8 &  \xmark &  \xmark & \cmark & CC-NC \\
        M4Singer \cite{zhang2022m4singer} & 2022 & CN & 700 & 29.8 & 20 & 1.5 &  \cmark & \cmark & \cmark & CC-NC \\
        SingStyle111 \cite{dai2023singstyle111} & 2023 & CN/EN/IT & 224 & 12.8 & 8 & 1.6 &   \cmark &  \xmark & \cmark & Restricted \\
        \midrule
        ACE-Opencpop & 2024 & CN & 100 & 128.9 & 30 & 4.3 & \cmark & \cmark & \cmark & CC-NC \\
        ACE-KiSing & 2024 &  CN/EN & 23 & 32.5 & 34 & 1.0 & \cmark & \cmark & \cmark & CC-NC \\
        \bottomrule
    \end{tabular}
    }
\vspace{-10pt}
    \label{tab:dataset}
\end{table*}